\documentclass[aps,prl,onecolumn,superscriptaddress,groupedaddress]{revtex4}  
\usepackage{graphicx}  
\usepackage{color, soul}
\usepackage{xcolor}
\usepackage{dcolumn}   
\usepackage{bm}        
\usepackage{amssymb}   
\usepackage{changepage}
\usepackage{amsmath}
\usepackage{hyperref}
\usepackage{float}
\DeclareUnicodeCharacter{00A0}{ }
\hypersetup{
  colorlinks=false,
  linkbordercolor=blue
}

\usepackage[a4paper,bindingoffset=0.1in,%
            left=1.5cm,right=1.5cm,top=2cm,bottom=2.5cm,%
            footskip=0.5in]{geometry}
\usepackage{blindtext}

\begin{document}


\newlength{\halfpagewidth}
\setlength{\halfpagewidth}{\linewidth}
\divide\halfpagewidth by 2
\newcommand{\leftsep}{%
\noindent\raisebox{4mm}[0ex][0ex]{%
\makebox[\halfpagewidth]{\hrulefill}\hbox{\vrule height 3pt}}%
}
\newcommand{\rightsep}{%
\noindent\hspace*{\halfpagewidth}%
\rlap{\raisebox{-3pt}[0ex][0ex]{\hbox{\vrule height 3pt}}}%
\makebox[\halfpagewidth]{\hrulefill} } 
    
\title{How Thermal Annealing Process Determines the Inherent Structure Evolution in Amorphous Silicon: An Investigation from Atomistic Time Scales to Experimental Time Scales}
\author{Yanguang~Zhou}
\email{maeygzhou@ust.hk}
\affiliation{Department of Mechanical and Aerospace Engineering, The Hong Kong University of Science and Technology, Clear Water Bay, Kowloon, Hong Kong}

\date{\today}

\begin{abstract}
\begin{center}
\textbf{Abstract}
\end{center}
The annealing treatment in the advanced manufacturing process, e.g., laser assisted manufacturing, determines the final state of glasses which is critical to its thermal, electrical and mechanical properties. Energy barriers analysis based on the potential energy surface offers an effective way to study the microscopic evolution of the inherent structures during the annealing process in a broadening timescale range, i.e., from atomistic timescale ($\sim$ ps) to experimental timescale ($\sim$  s). Here, we find the distribution activation energy barriers in the potential energy surface can be divided into three regimes: 1, the distribution mainly follows the Rayleigh distribution when the annealing rate $\dot{R}<5\times {{10}^{15}}\ \text {K/s}$; 2, two different modes, i.e., an exponentially decaying mode and a Rayleigh distribution mode, are found in the spectra when the annealing rate $5\times {{10}^{15}}\ \text{K/s}<\dot{R}<instant$; 3, the spectra is almost following the exponentially decaying mode when the system is under the instant annealing process. However, the spectra of relaxation energy barriers only show an exponentially decaying mode with a decreasing decay parameter. A multi timescale model for any specific annealing rate, which is beyond the limit of the conventional atomistic simulations, i.e., molecular dynamics simulations, is then proposed based on the distribution of the energy barriers. Such a model enables quantitative explanations and predictions of the heat release during the annealing process of the nanocalorimetry measurements or laser assisted manufacturing.
\end{abstract}
\setulcolor{blue}
\maketitle
\twocolumngrid 

Structural defects such as dislocations and grain boundaries in crystal cannot be well characterized in amorphous material due to its topology of atomic connectivity varies locally. For such a reason, amorphous material, e.g., amorphous silicon, has many promising physical, chemical and mechanical properties comparing to their crystalline counterparts, e.g., extremely low thermal conductivity \cite{Goldsmid1983, Larkin2014, Kwon2016, He2011, Liu2009, Moon2018, Saaskilahti2016, Zhou2016, Zhou2015, Zhou20162}, exceptional charge capacity \cite{Ding2015, Ding2017} and high strength \cite{Patinet2016, Fan2013}. In amorphous material, many properties are strongly related to the degrees of relaxation in the system which can be quantified via the inherent structural energy (ISE), i.e., $E_\text{IS}$, calculating by removing the kinetic energy \cite{Kallel2010}. Therefore, the properties such as damage evolution \cite{Patinet2016, Fan2015} and flow stress \cite{Fan20132} of amorphous materials can be modulated through ISE. One of most popular approach to vary the ISE of amorphous materials is to adjust the dynamic annealing process \cite{Liu2020, Fan2017, Ramachandramoorthy2016}. Recent development in ultrafast laser pulse techniques \cite{Shou2019} make the annealing rate as high as 10$^6$ K/s which is still an unresolved challenge to date to the traditional modelling tools, i.e., molecular dynamics (MD) techniques \cite{Fan2017, Deringer2018, Fan2014, Chowdhury2016}. The potential energy surface (PES) generated based on the inherent structure energy has been proved to be capable of characterizing the complex phenomenology in amorphous materials \cite{Fan2015, Liu2020, Fan2017, Fan2014, Yan2016}, and more importantly they allow focusing on the energy distribution rather than the dynamic process, making it possible to capture the experimental timescale annealing rates, i.e., $10^{-2}$ $\sim$ $10^6$ K/s \cite{Shou2019, Wan2011, Yuan2017}.

\begin{figure}[H]
\hspace*{-4mm} 
\setlength{\abovecaptionskip}{0.05in}
\setlength{\belowcaptionskip}{-0.1in}
\includegraphics [width=3.5in]{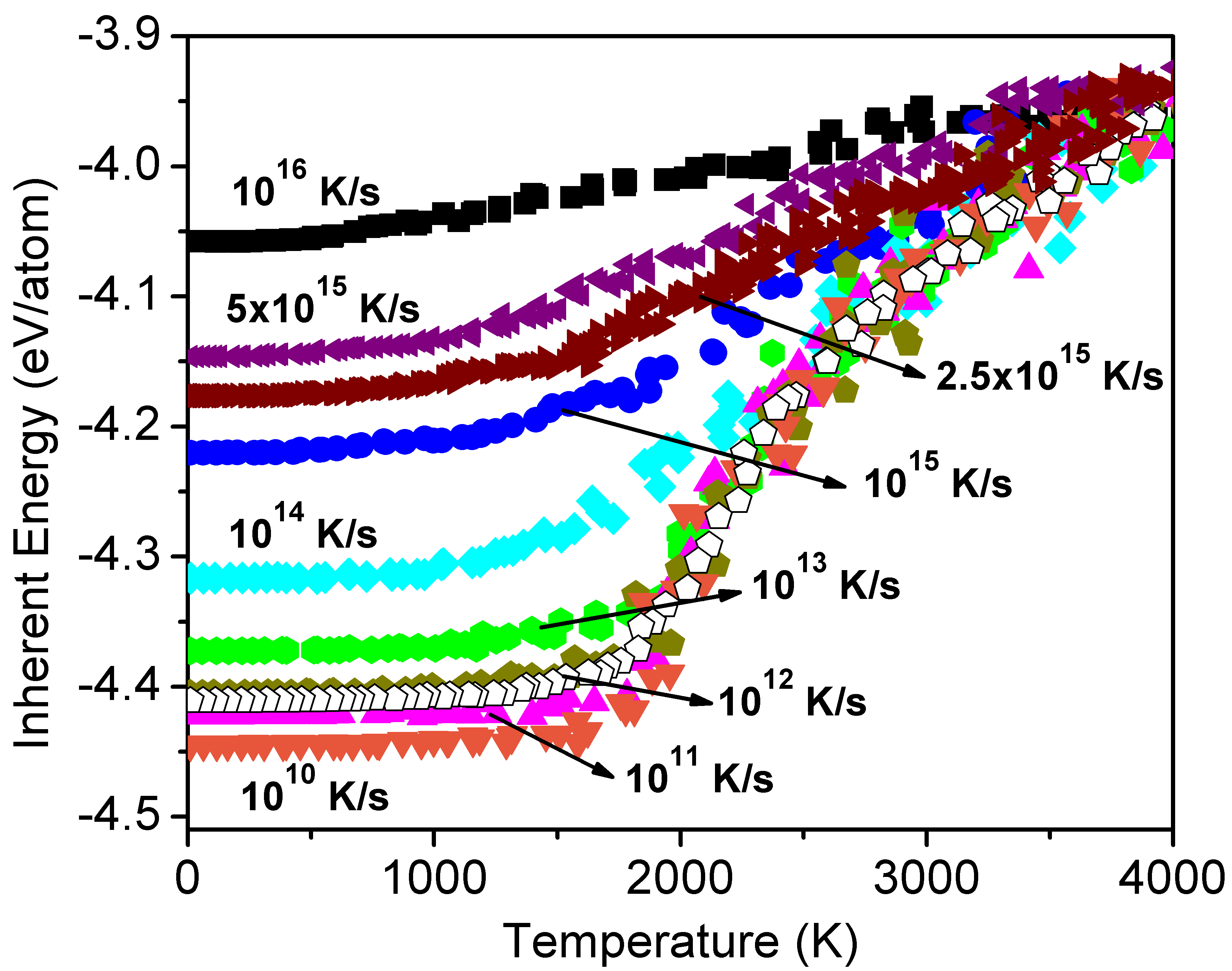}
\caption{The inherent energy of the systems during the annealing process with different annealing rates, i.e., 10$^{10}$ K/s up to 10$^{16}$ K/s, in molecular dynamics simulations. The solid dots are results from the models with 1728 atoms, and the open dots are calculated by the system of 4096 atoms. Our results show the system with 1728 atoms is large enough to ignore the size effects.}
\label{fig:F1}
\end{figure}

To generate the PES, the elementary processes are the hopping between neighboring local minima. This process consists of two stages: 1, the activation stage which is from the initial state to the connecting saddle state with activation barrier ${{E}_{\text{A}}}$; and 2, the relaxation stage, i.e., from the saddle state to the final state with energy barrier ${{E}_{\text{R}}}$. Then, the time evolution of ${{E}_\text{IS}}$ can be expressed in the form of \cite{Liu2020, Fan2017, Derlet2013}
\begin{equation}
\begin{aligned}
 \frac{d{{{\bar{E}}}_\text{IS}}}{dt}= & \ \frac{dT}{dt}\frac{d{{{\bar{E}}}_\text{IS}}}{dT}
 \approx \dot{R}\left[ \frac{{{{\bar{E}}}_\text{IS}}(T+\Delta T)-{{{\bar{E}}}_\text{IS}}(T)}{\Delta T} \right] \\ 
 =&f\cdot \exp \left( -\frac{{{{\bar{E}}}_\text{A}}}{{{k}_{b}}T} \right)\left( {{{\bar{E}}}_\text{A}}-{{{\bar{E}}}_\text{R}} \right) \\ 
\end{aligned}
\label{eqn:A1}  
\end{equation}
in which, $f$ is the jump frequency which includes the entropy effects \cite{Goldstein1976, Johari1977, Berthier2011}, ${{k}_{b}}$ is the Boltzmann constant,   is the temperature of the current inherent structure, $\dot{R}$ is the annealing rate, and ${{\left[ f\cdot \exp \left( -{{{\bar{E}}}_\text{A}}/{{k}_{b}}T \right) \right]}^{-1}}$ represents the average residence time in current inherent structure. ${{\bar{E}}_\text{A}}$ and ${{\bar{E}}_\text{R}}$ are the average activation and relaxation energy barriers, which consider the effect of multiple transition pathways in the PES of amorphous materials, and can be calculated via \cite{Fan2017, Derlet2013}
\begin{small}
\begin{equation}
{{\bar{E}}_\text{A}}=-{{k}_{b}}T\cdot \ln \left[ \int{P\left( {{E}_\text{A}}|{{E}_\text{IS}} \right){{e}^{-\frac{{{E}_\text{A}}}{{{k}_{b}}T}}}d{{E}_\text{A}}} \right]
\label{eqn:A2}  
\end{equation}
\end{small}
\begin{small}
\begin{equation}
{{\bar{E}}_\text{R}}=\int{{{E}_\text{R}}\cdot P\left( {{E}_\text{R}}|{{E}_\text{IS}} \right)d{{E}_\text{R}}}
\label{eqn:A3}  
\end{equation}
\end{small}

\noindent
where $P\left( {{E}_{A}}|{{E}_{ISE}} \right)$ and $P\left( {{E}_{R}}|{{E}_{ISE}} \right)$ are the activation and relaxation energy spectra which are calculated via the activation-relaxation technique (ART) \cite{Cancs2009, Barkema1996} – a method of atomistic simulation known to be able to capturing the saddle point states and providing the standard PES samples including these high activation energy barrier events that cannot be accessible by MD simulations in amorphous materials. Here, the crystalline silicon with 1728 atoms with interaction depicting using Tersoff potential \cite{Tersoff1989} is first equilibrated at a high temperature (4000 K) liquid state, and then cooled to 10 K with nine annealing rates ranging from 10$^{10}$ K/s up to 10$^{16}$ K/s and the instant quench using MD simulations. The size of the system is 3.26 nm$\times$3.26 nm$\times$3.26 nm, with periodic boundary conditions applied to all the three directions. The system with 4096 atoms is used to validate the size effect in our results can be ignored. \textbf{Figure ~\ref{fig:F1}} shows the evolution of the ISE at various temperatures in MD simulations at the annealing rate of 10$^{10}$ K/s up to 10$^{16}$ K/s. In the high annealing rate process, e.g., $\dot{R}={{10}^{16}}\ \text{K/s}$, the system would jump out of the equilibrium state and eventually turn into the glassy state with an almost constant ISE (black dots in \textbf{Figure ~\ref{fig:F1}}) since the timescale of cooling rate cannot keep up with the intrinsic timescale in supercooled liquid. With the decreasing of the annealing rate, e.g., $\dot{R}<{{10}^{16}}\ \text{K/s}$,  the systems at temperature above 1800 K can reach equilibrium state easily and their corresponding ISEs decrease quickly, and then froze to glassy states with different ISEs with temperature decreasing below 1800 K. Next, the ISs generated in the MD runs are going to be used as input for ART simulations to calculate the ${{E}_\text{A}}$ and ${{E}_\text{R}}$ spectra. For each of the ten input structures, 5000 ART searchers with different random perturbation directions are employed. The magnitude of perturbation displacement is fixed at 0.5 $\text{\AA}$, and the system is relaxed to the saddle point following the Lanczos algorithm \cite{Barkema1996, Malek2000} when the curvature of the PES is less than -0.01 eV$\cdot $\AA$^{-2}$ and the force of the system is less than 0.05 eV$\cdot $\AA$^{-1}$. After removing the failed and redundant searches, around 3,000 different searches are left for each IS.            

\begin{figure*}
\setlength{\abovecaptionskip}{0.1in}
\setlength{\belowcaptionskip}{-0.1in}
\includegraphics [width=6.5in]{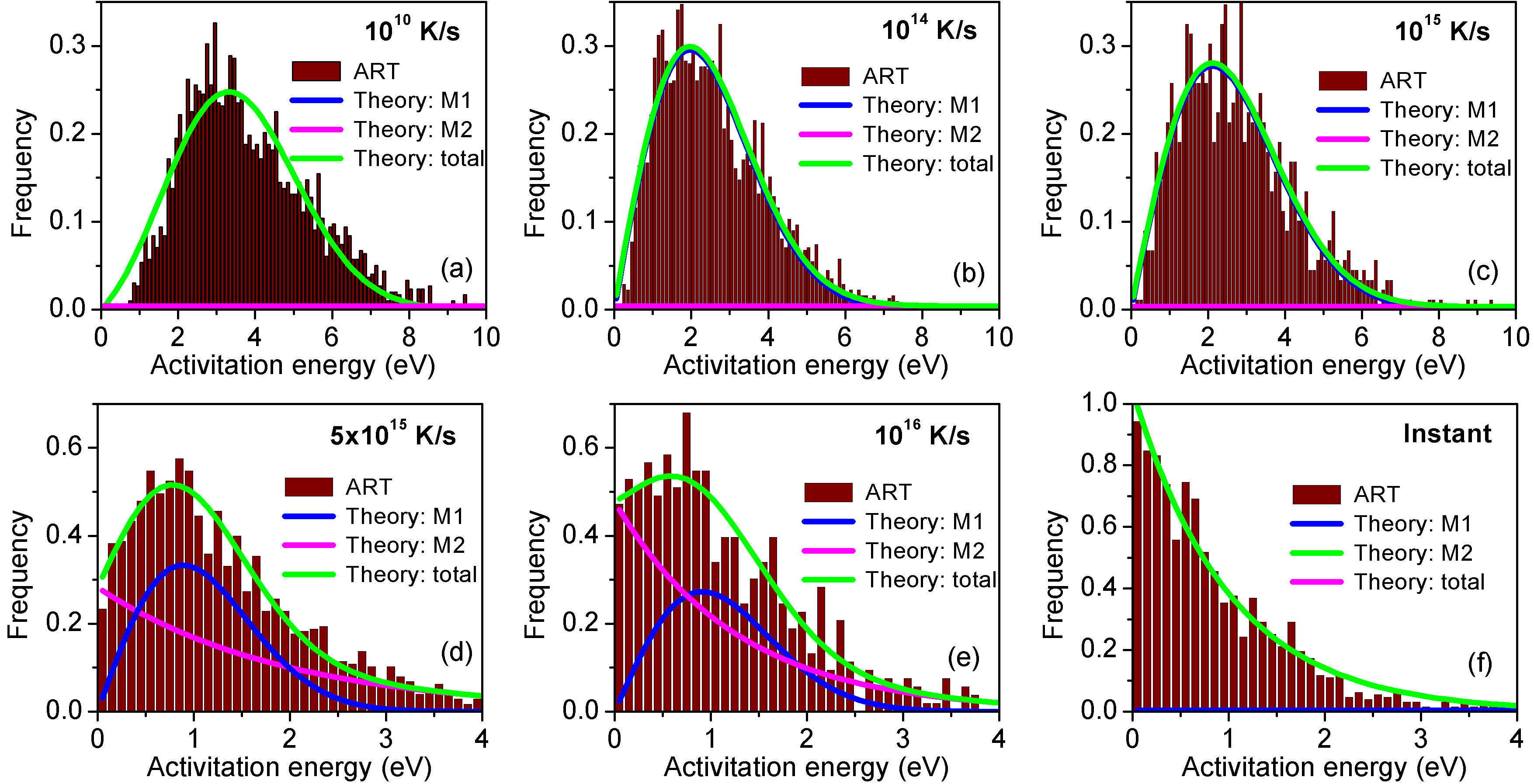}
\caption{The raw histogram data on the activation energy barrier spectra in different samples generated at various annealing rates (a) $10^{10}$ K/s, (b) $10^{14}$ K/s, (c) $10^{15}$ K/s, (d) 5$\times$$10^{15}$ K/s, (e) 10$^{16}$ K/s and (f) instant quench obtained by ART. The spectra can be fitted into two modes: an Rayleigh distribution (M1 mode, blue lines in the Figure) with an obvious peak and an exponential decay mode (M2 mode, purple lines in the Figure), using Eq.\ (\ref{eqn:A4}).}
\label{fig:F2}
\end{figure*}
\begin{figure}
\hspace*{-4mm} 
\setlength{\abovecaptionskip}{0.1in}
\setlength{\belowcaptionskip}{-0.2in}
\includegraphics [width=3.5in]{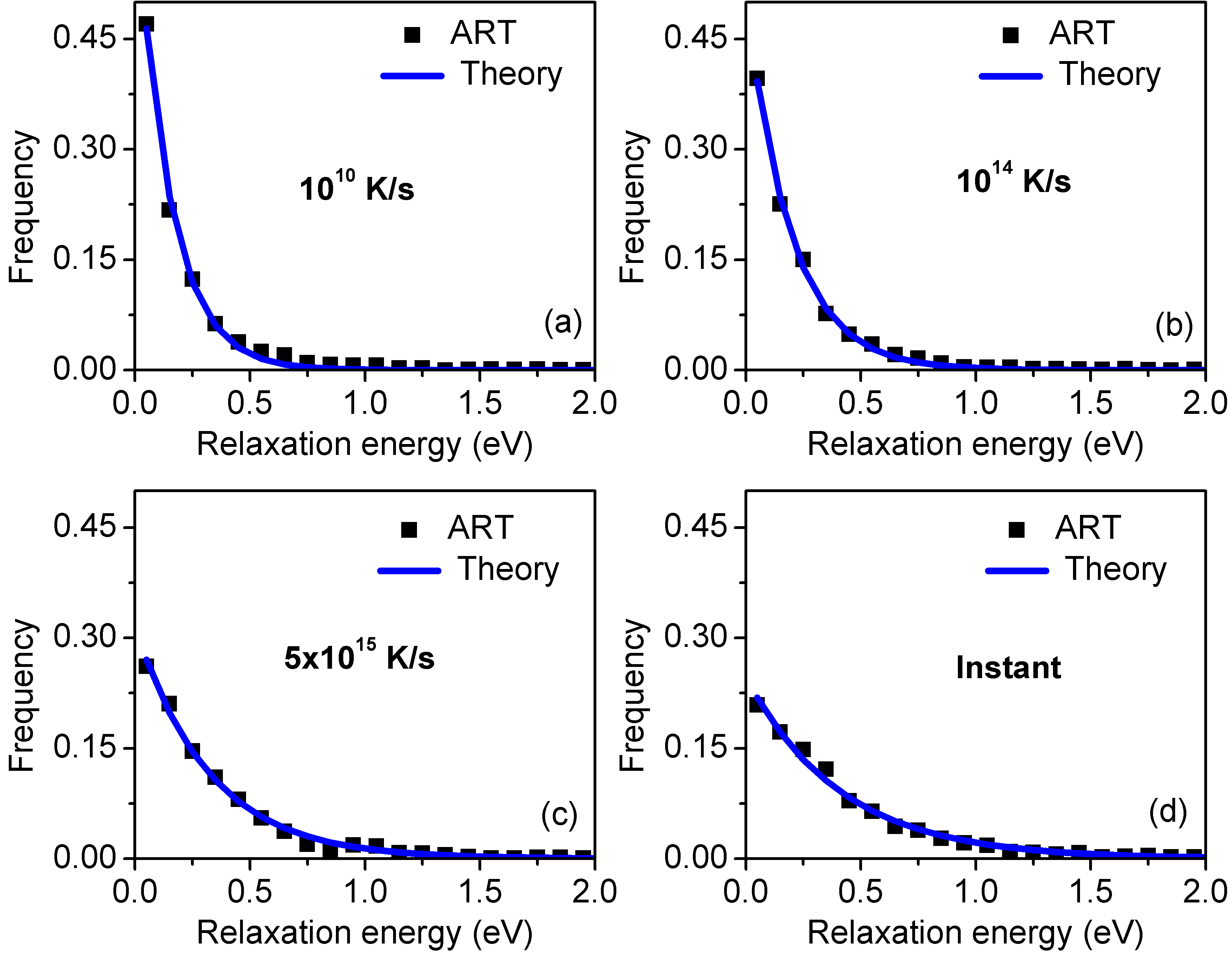}
\caption{The raw data on the relaxation energy barrier spectra in different samples generated at various annealing rates (a)  10$^{10}$ K/s, (b)  10$^{14}$ K/s, (c) 5$\times$10$^{10}$ K/s and (d) instant quench obtained by ART. Lines are fitted using Eq.\ (\ref{eqn:A5}).}
\label{fig:F3}
\end{figure}

We then focus on the activation energy spectra $P\left( {{E}_\text{A}}|{{E}_\text{IS}} \right)$ and relaxation energy spectra $P\left( {{E}_\text{R}}|{{E}_\text{IS}} \right)$. \textbf{Figure ~\ref{fig:F2}} shows the activation energy spectra obtained for these samples. It is not surprising to find faster annealing rate yield smaller activation energy range since the intrinsic timescale is larger than the timescale of cooling rate and therefore it is difficult for the system to catch up to the high saddle energy states \cite{Fan2017}. More interesting, it is found that $P\left( {{E}_\text{A}}|{{E}_\text{IS}} \right)$ can be divided into three regimes: 1, the spectra dominantly follows the Rayleigh distribution with a clear peak when $\dot{R}<5\times {{10}^{15}}\ \text{K/s}$ (M1 mode in our manuscript, \textbf{Figure ~\ref{fig:F2}a-c}); 2, the spectra can be decomposed into two different modes: an exponentially decaying mode (M2 mode) and a Rayleigh distribution mode (M1 mode) when $5\times {{10}^{15}}\ \text{K/s}<\dot{R}<instant$ \textbf{Figure ~\ref{fig:F2}d-e}); 3, the spectra is mainly following the M2 mode when the system is under the instant annealing process (\textbf{Figure ~\ref{fig:F2}f}). Based on the probability distribution of the spectra, the spectra can be represented phenomenologically using \cite{Liu2020, Fan2017}:
\begin{equation}
\begin{aligned}
P\left( {{E}_{\text{A}}}|{{E}_{\text{IS}}} \right)= \ & {{P}_\text{{M}1}}\left( {{E}_{\text{A}}}|{{E}_{\text{IS}}} \right)+{{P}_\text{{M}2}}\left( {{E}_\text{{A}}}|{{E}_\text{{IS}}} \right) \\ 
= \ & {{W}_{1}}\cdot {{E}_\text{{A}}}\cdot \exp \left[ -\frac{{{({{E}_\text{{A}}}-\mu ({{E}_\text{{IS}}}))}^{2}}}{2{{\sigma }^{2}}} \right]+\\ 
 \ & {{W}_{2}}\cdot \frac{1}{{{\xi }_\text{{A}}}}\cdot \exp \left( -\frac{{{E}_\text{{A}}}}{{{\xi }_\text{{A}}}} \right) \\ 
\end{aligned}
\label{eqn:A4}  
\end{equation}
where ${{W}_{1}}$ and ${{W}_{2}}$ are the weigh factors related to the amplitude of M1 and M2 modes, respectively. $\mu $ is the location parameter of M2, $\sigma $ is the width of the M1 mode, and ${{\xi }_{\text{A}}}$ is the decay parameter of M2 mode. It is worth noting the integral of $P\left( {{E}_\text{A}}|{{E}_\text{IS}} \right)$ should be equal to one due to the normalization condition of the spectra. As seen in \textbf{Figure ~\ref{fig:F2}}, when the inherent structure energy becomes higher, i.e., the higher annealing rate, the amplitude of M1 becomes smaller and disappear eventually (\textbf{Figure ~\ref{fig:F2}}), while the amplitude of M2 shifts to a larger value and becomes the main mode when the system is under instant annealing (\textbf{Figure ~\ref{fig:F2}f}). In the similar way, we find the spectra of relaxation energy $P\left( {{E}_\text{R}}|{{E}_\text{IS}} \right)$ can be fitted in the phenomenological formula of \cite{Kallel2010, Liu2020, Fan2017, Derlet2013}
\begin{equation}
P\left( {{E}_{\text{R}}}|{{E}_{\text{IS}}} \right)=\frac{1}{{{\xi }_{\text{R}}}}\cdot \exp \left( -\frac{{{E}_{\text{R}}}}{{{\xi }_{\text{R}}}} \right)
\label{eqn:A5}  
\end{equation}
in which, ${{\xi }_\text{R}}$ is the decay parameter. Unlike the previous studies found the relaxation spectra is independent of the annealing rate \cite{Kallel2010}, our results show that higher annealing rate will lead to a broad range of the relaxation energy barrier (\textbf{Figure ~\ref{fig:F3}}). As discussed in Ref. \cite{Fan2015, Fan2014}, the higher annealing rate will lead to a faster exponential decay behavior of relaxation energy barrier distribution indicating a localized relaxation process which is found in various systems, and with decreasing of the annealing rate, the probability of the relaxation energy barrier distribution will become broader due to the combination of localization and cascade (i.e., self-organized criticality which has been observed universally in many different systems experimentally \cite{Krisponeit2014} and numerically \cite{Salerno2012}) relaxations of ${{E}_\text{R}}$ in the PES. 

\begin{figure*}
\setlength{\abovecaptionskip}{0.1in}
\setlength{\belowcaptionskip}{-0.1in}
\includegraphics [width=6.5in]{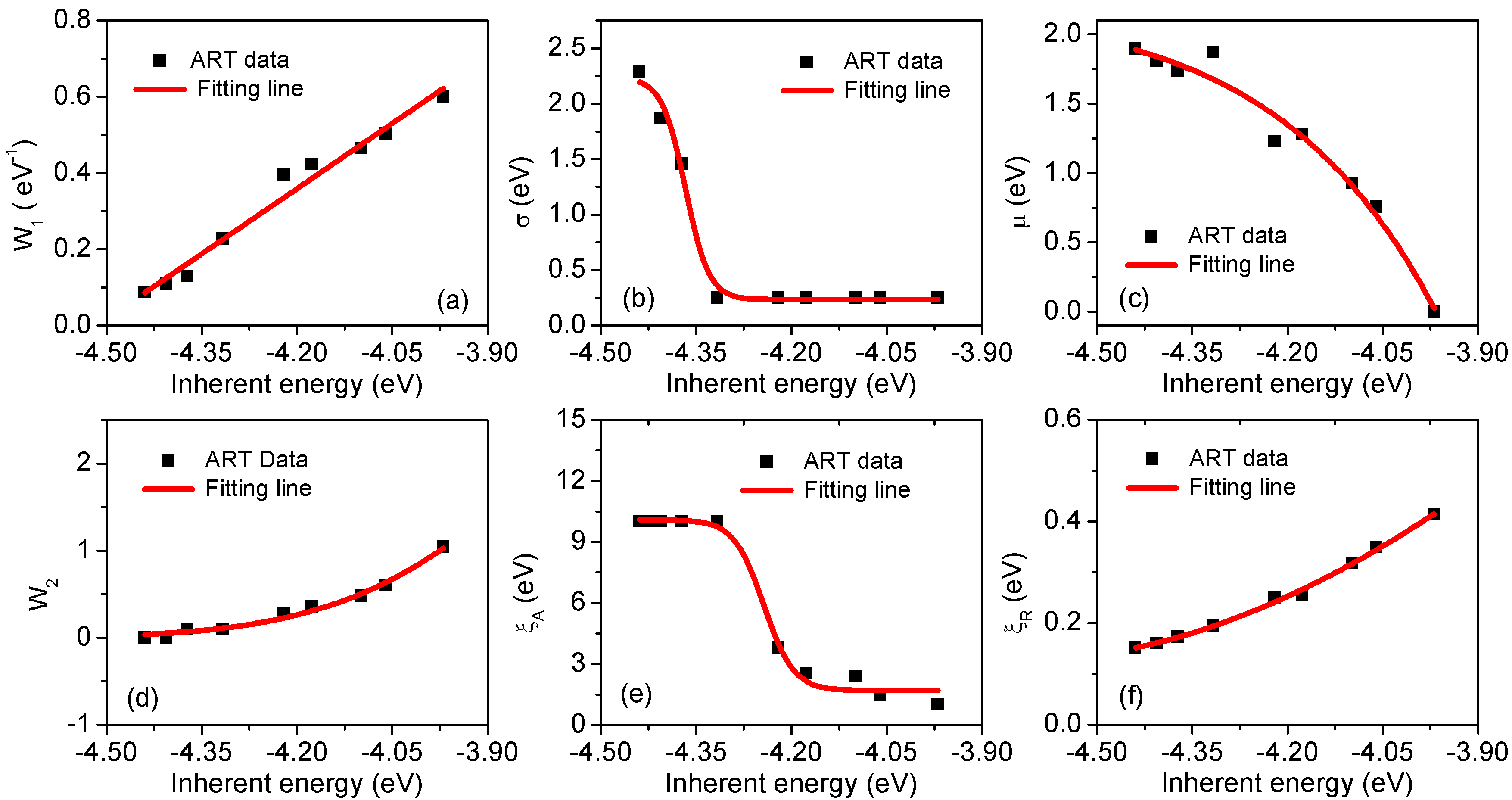}
\caption{The magnitude of (a) ${{W}_{1}}$, (b) $\sigma$, (c) $\mu$, (d) ${{W}_{2}}$, (e) ${{\xi }_{\text{A}}}$ in Eq.\ (\ref{eqn:A4}) and (f) ${{\xi }_{\text{R}}}$ in Eq.\ (\ref{eqn:A5}) v.s. the inherent energy $E_{\text{IS}}$. To make Eq.\ (\ref{eqn:A1})  to be used at any specific annealing rate, we fit the discrete data using empirical equations: (a) ${{W}_{1}}=-92.4+97.6\cdot \exp ({{E}_{\text{IS}}}/81.42)$, (b) $\sigma=0.233+(2.234-0.233)/\left[ 1+\exp (({{E}_{\text{IS}}}+4.367)/0.019) \right]$, (c) $\mu=2.224-17.25\times {{10}^{6}}\cdot \exp ({{E}_{\text{IS}}}/0.25)$, (d) ${{W}_{2}}=10\cdot \exp (-48.61-28.30{{E}_{\text{IS}}}-4.19E_{\text{IS}}^{2})$, (e) ${{\xi }_{\text{A}}}=1.308+1.370\times {{10}^{-28}}\cdot \exp (-{{E}_{\text{IS}}}/0.066)+1.242\times {{10}^{-28}}\cdot \exp (-{{E}_{\text{IS}}}/0.065)$ and (f) ${{\xi }_{R}}=3.836-1.075{{E}_{\text{IS}}}-0.985E_{\text{IS}}^{2}-0.125E_{\text{IS}}^{3}$.}
\label{fig:F4}
\end{figure*}

To solve Eq.\ (\ref{eqn:A1}) at any given ISE, we need prepare the forms of ${{W}_{1}}$, $\mu$, $\sigma$, ${{W}_{2}}$, ${{\xi }_{\text{A}}}$ and ${{\xi }_{\text{R}}}$. In our study, we have nine samples which can provide nine data of each parameter mentioned above. What is worth noting is that a wider range of ISE, particularly those ISEs at slower annealing rate, e.g., $\dot{R}<{{10}^{10}}\ \text{K/s}$, can make sure the fitting process of the parameters more accurately. But it will take prohibiting computing time or even out of the power of MD simulations. In this study, we firstly fit these parameters using appropriate functions and then compare their generated average saddle-initial or final minimal energy differences with MD results directly. Based on the data generated from the ART results, we can find ${{W}_{1}}=\ -92.4+97.6\cdot \exp ({{E}_{\text{IS}}}/81.42)$ (\textbf{Figure ~\ref{fig:F4}a}), $\sigma\ =\ 0.233+(2.234-0.233)/\left[ 1+\exp (({{E}_{\text{IS}}}+4.367)/0.019) \right]$ (\textbf{Figure ~\ref{fig:F4}b}), $\mu =2.224-17.25\times {{10}^{6}}\cdot \exp ({{E}_{\text{IS}}}/0.25)$ (\textbf{Figure ~\ref{fig:F4}c}), ${{W}_{2}}=10\cdot \exp (-48.61-28.30{{E}_{\text{IS}}}-4.19E_{\text{IS}}^{2})$ (\textbf{Figure ~\ref{fig:F4}d}), ${{\xi }_{\text{A}}}=1.308+1.370\times {{10}^{-28}}\cdot \exp (-{{E}_{\text{IS}}}/0.066)+1.242\times {{10}^{-28}}\cdot \exp (-{{E}_{\text{IS}}}/0.065)$ (\textbf{Figure ~\ref{fig:F4}e}), ${{\xi }_{R}}=3.836-1.075{{E}_{\text{IS}}}-0.985E_{\text{IS}}^{2}-0.125E_{\text{IS}}^{3}$ (\textbf{Figure ~\ref{fig:F4}f}). The average saddle-initial or final energy difference is then calculated for validating our fitting formulas via     
\begin{equation}
\begin{aligned}
&{{\bar{E}}_{{saddle}-initial\ or\ final}}=\\ 
&\int{{{E}_{\text{A}\ or\ \text{R}}}\cdot P\left( {{E}_{\text{A}\ or\ \text{R}}}|{{E}_{\text{IS}}} \right)d{{E}_{\text{A}\ or\ \text{R}}}}
\end{aligned}
\label{eqn:A6}  
\end{equation}

\begin{figure}[H]
\hspace*{-6mm} 
\setlength{\abovecaptionskip}{0.1in}
\setlength{\belowcaptionskip}{-0.1in}
\includegraphics [width=3.5in]{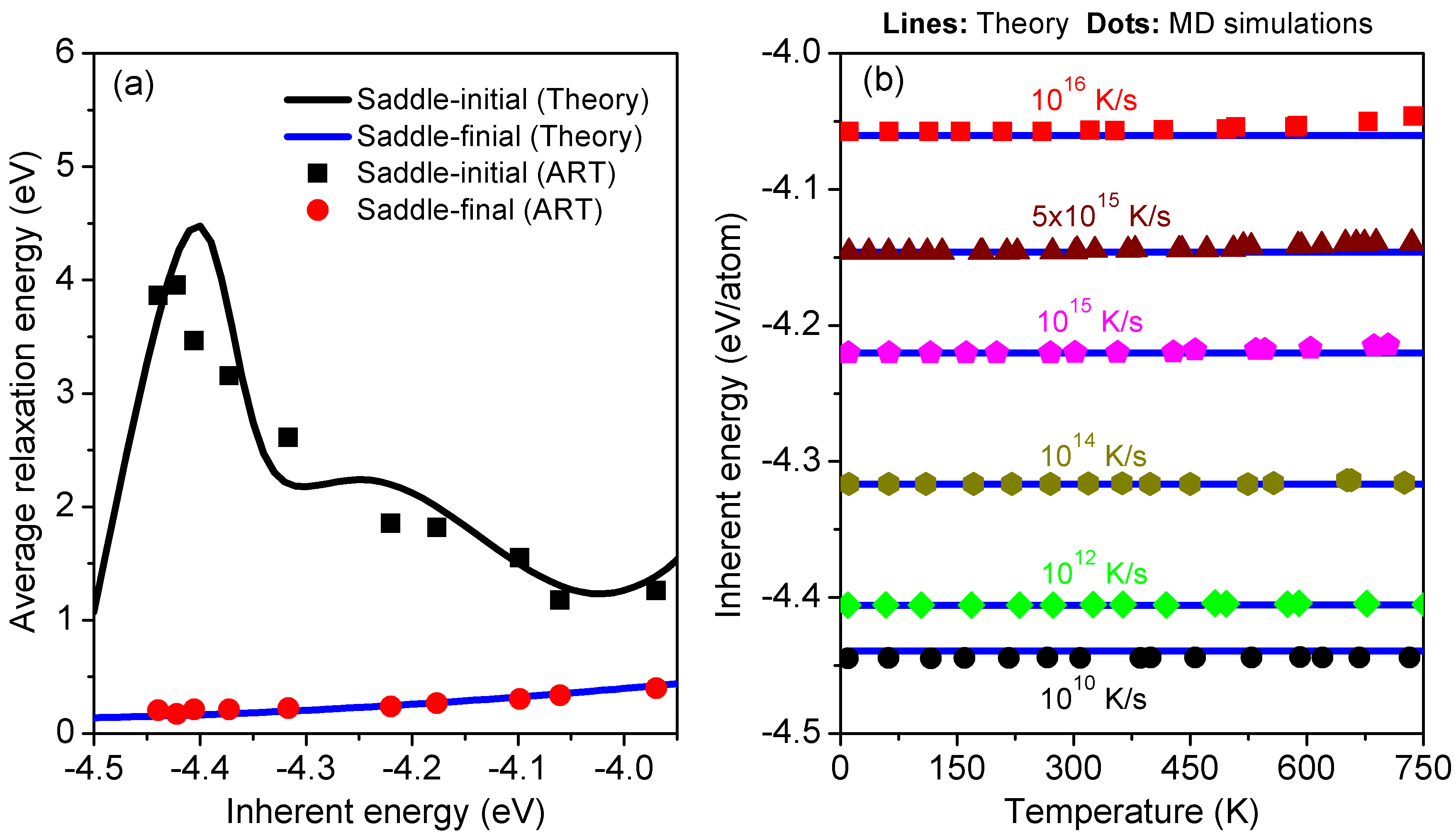}
\caption{(a) The average relaxation energy calculated using ART data and empirical formulas fitted in \textbf{Figure ~\ref{fig:F4}}. (b) The inherent energy changes with temperature at various annealing rates, dots are from MD simulations and lines are calculated using Eq.\ (\ref{eqn:A1}).}
\label{fig:F5}
\end{figure}

The lines in \textbf{Figure ~\ref{fig:F5}a} are calculated using Eq.\ (\ref{eqn:A6}) with the functions of the parameters fitted above (\textbf{Figure ~\ref{fig:F4}}), and the dots in Figure 5a are the results obtained from discrete ART data. The good agreement between the lines and the dots in Figure 5a indicates rationality of the fitting forms of ${{W}_{1}}$, $\mu$, $\sigma$, ${{W}_{2}}$, ${{\xi }_{\text{A}}}$ and ${{\xi }_{\text{R}}}$. To fully obtain the dynamic process of the ISE during the annealing or heating process at arbitrary annealing rate, or equivalently solving Eq. (1), the hypothesis on the jump frequency is still necessary. As firstly proposed by Goldstein \cite{Goldstein1976} and Johari \cite{Johari1977}, the jump frequency $f$ depends on the entropy of the system which is related to the temperature and ISE. Since the ISE is a function of temperature as well, we then assume $f$ only depends on the temperature and can be expressed as ${{f}_{0}}\exp[-1.25\times{{10}^{-4}}(T-300)]/{{(9.6T)}^{2}}$ \cite{Fan2017} where ${{f}_{0}}$ is the zero temperature jump frequency and is set as $1\times {{10}^{9}}$ THz \cite{Fan2013, Yan2016, Fan2012}. Next, the Eq.\ (\ref{eqn:A1}) can be solved at arbitrary annealing rate $\dot{R}$, and good agreements between our theoretical predictions with MD simulations are found (\textbf{Figure ~\ref{fig:F5}b}).

The model developed above is then used to study the heat release of the system $Q$ at experimental annealing rates, i.e., ${{10}^{-2}}\sim{{10}^{6}}\ \text{K/s}$ \cite{Shou2019, Wan2011, Yuan2017}, which can be written as:
\begin{equation}
\begin{aligned}
Q=&\frac{d\bar{E}}{dT}=\frac{d{{{\bar{E}}}_\text{IS}}+d{{{\bar{E}}}_{kinetic}}}{dT}\\ 
\approx &\left[ \frac{{{{\bar{E}}}_\text{IS}}(T+\Delta T)-{{{\bar{E}}}_\text{IS}}(T)}{\Delta T}+1.5{{k}_{b}} \right]
\end{aligned}
\label{eqn:A7}  
\end{equation}
Combining Eqs.\ (\ref{eqn:A1}) and \ (\ref{eqn:A7}), we can know
\begin{equation}
\begin{aligned}
Q=\frac{f}{{\dot{R}}}\cdot \exp \left( -\frac{{{{\bar{E}}}_\text{A}}}{{{k}_{b}}T} \right)\left( {{{\bar{E}}}_\text{A}}-{{{\bar{E}}}_\text{R}} \right)+1.5{{k}_{b}}
\end{aligned}
\label{eqn:A8}  
\end{equation}

\vspace*{-4mm} 

\begin{figure}
\hspace*{-6mm} 
\setlength{\abovecaptionskip}{0.1in}
\setlength{\belowcaptionskip}{-0.1in}
\includegraphics [width=3.5in]{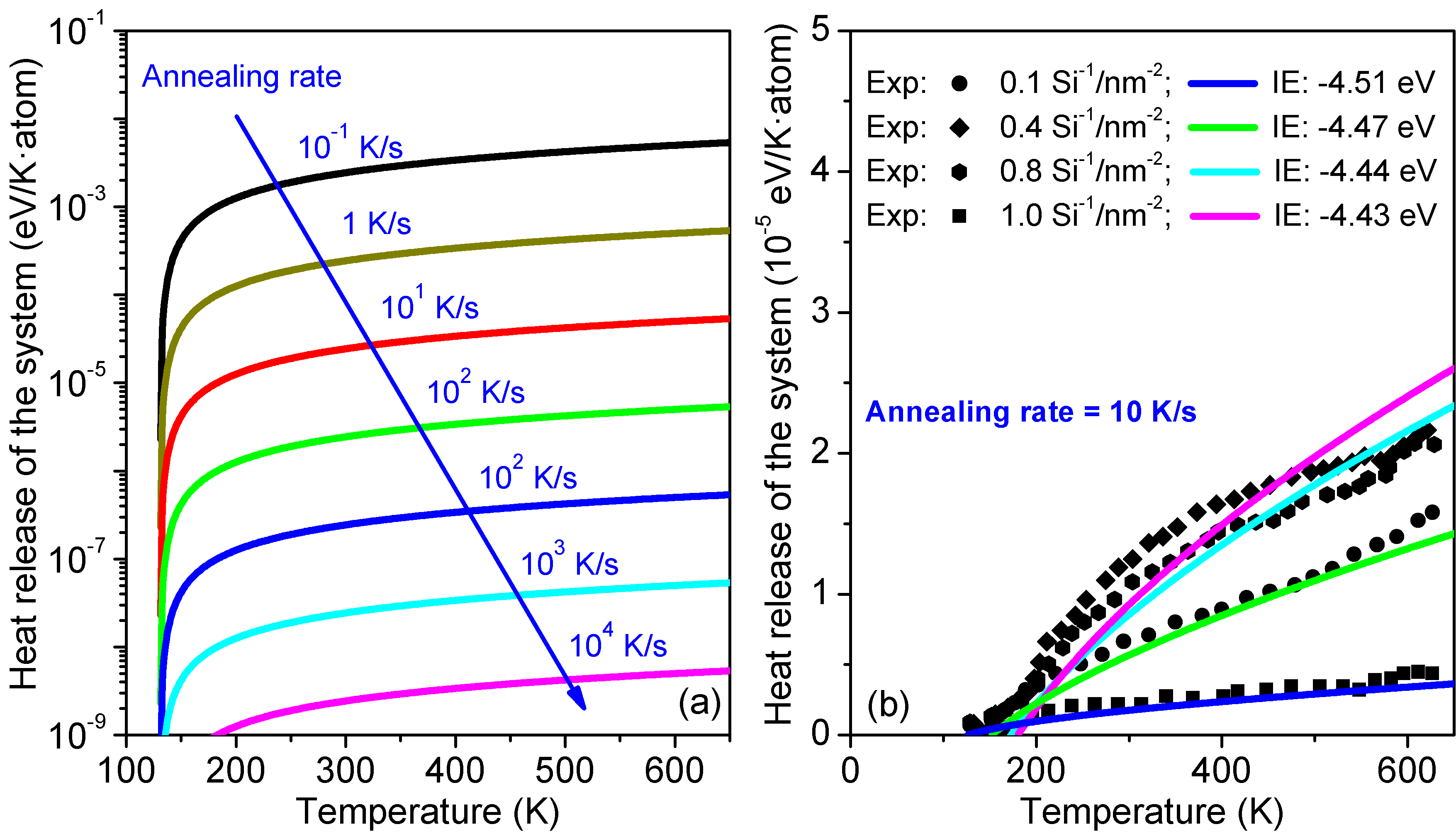}
\caption{(a) the predicted heat release during the annealing process with different annealing rates using Eq.\ (\ref{eqn:A8}), which we assume the ${{E}_{\text{IS}}}=-4.51\ \text{eV}$. (b) The het release when the annealing rate is 10 K/s   with appropriate inherent energy, experimental measurements are from Ref. \cite{Karmouch2007}, excellent agreement between theoretical predictions and experimental data is found.}
\label{fig:F6}
\end{figure}

\textbf{Figure ~\ref{fig:F6}a} shows the heat release of the system during the annealing process of amorphous Si at the rates of ${{10}^{-2}}\ \sim {{10}^{6}}\ \text{K/s}$ using Eq.\ (\ref{eqn:A8}). At high temperature range ($\geq$ 200 K), the systems can be equilibrated at very short time scale and therefore the abundant heat will be released. However, at the low temperature range ($\leq$ 200 K), the annealing rate is too fast to equilibrate the system, and the system is undergoing a normal ageing behavior \cite{Liu2020, Fan2017}. We now try to compare our predictions with the experimental measurements \cite{Karmouch2007}. In the experiments, the Si ions are implanted into the amorphous samples, and then the released heat of the system is measured. By comparing the results in Figure 3 and 4 in Ref. \cite{Karmouch2007}, we can assume the number of atoms of the ion-implanted region is about 10$^{16}$. By choosing the ISE of the systems and annealing rate carefully, our theoretical predictions agree very well with the experimental measurement (\textbf{Figure ~\ref{fig:F6}b}). It is worth noting that the higher density implanted ions will cause higher ISE of the system as we assumed here. Meanwhile, the annealing rate can be as high as ${{10}^{6}}\ \text{ K/s}$ in the laser flash experiments \cite{Shou2019} and as low as ${{10}^{-2}}\ \text{K/s}$ in the  furnace lamp annealing experiments \cite{Shrestha2018}, here we use the medium annealing rate 10 K/s in our model.

We also note that in Ref. \cite{Kallel2010}, the potential energy surface of activation and relaxation energy barriers in amorphous Si is almost independent on the inherent structures. The reason for this may be due to the original structures used in their study are the same, and the inherent energy is varying via deleting atoms in the system. In our paper, we generate the inherent structures via mimicing the real experimental annealing process, we find the distribution of activation energy can be decomposed in two modes which is found in other systems such as amorphous Cu$_{56}$Zr$_{44}$ \cite{Fan2017} and the distribution of relaxation energy is varying with the annealing rate which is also observed in other systems, e.g., Stillinger-Weber silicon \cite{Middleton2001}. We would also like to emphasize that the accuracy of our theoretical predictions depends on the accuracy of $P\left( {{E}_\text{A}}|{{E}_\text{IS}} \right)$ and $P\left( {{E}_\text{R}}|{{E}_\text{IS}} \right)$ over the entire range of inherent structure energy. Molecular dynamics simulations can assure the parameters of $P\left( {{E}_\text{A}}|{{E}_\text{IS}} \right)$ and $P\left( {{E}_\text{R}}|{{E}_\text{IS}} \right)$ in high annealing rate from ${{10}^{10}}\ \text{ K/s}$ to ${{10}^{16}}\ \text{ K/s}$ and the instant annealing rate, i.e., inherent surface energy from -4.44 eV/atom to -3.97 eV/atom, in our theoretical model are reasonable, while there is no guarantee that their extrapolations are still suitable at experimental annealing rates, i.e., ${{10}^{-1}}\sim{{10}^{4}}\ \text{ K/s}$. However, the underlying mechanism behind the theoretical model proposed here is still valid since  Eq.\ (\ref{eqn:A8}) describes the physics process of hopping in potential energy surface.

In conclusion, the activation and relaxation energy barriers\rq distributions are extracted based on the extensive sampling of the potential energy surface of a-Si. We find the distribution of the activation energy barriers can be divided into three regimes: 1, the spectra dominantly follows the Rayleigh distribution when $\dot{R}<5\times {{10}^{15}}\ {K/s}$; 2, the spectra can be decomposed into two different modes: an exponentially decaying mode and a Rayleigh distribution mode when $5\times {{10}^{15}}\ {K/s}<\dot{R}<instant$; 3, the spectra is mainly following the exponentially decaying mode when the system is under the instant annealing process. The distribution of relaxation energy barrier is found to follow the exponentially decaying mode with a decreasing decaying parameter at the whole annealing rate range. Then, a long-time scale model beyond the limit of the conventional atomistic simulations used to capture the annealing process observed in experiments of generating a-Si is proposed based on distributions of the energy barriers. The heat release during the annealing process is calculated using our proposed long timescale model, and compared with the nanocalorimetry measurements. Excellent agreement between our numerical results and experimental measurements, makes sure the present model not only can unravel the microscopic information of evolution of the inherent structure but also applicable to many thermodynamics processes such as laser assisted manufacturing process.

Y.Z. thanks startup fund from Hong Kong University of Science and Technology (HKUST). Y. Z. gratefully acknowledges Mr. Zhitong Bai (University of Michigan-Ann Arbor) for valuable discussions.

\end{document}